\documentclass[11pt,a4paper]{article}
\usepackage[T1]{fontenc}
\usepackage[utf8]{inputenc}
\usepackage{lmodern}
\usepackage[left=1in,right=1in,top=0.85in,bottom=0.85in]{geometry}
\usepackage{amsmath,amssymb,amsthm,mathtools}
\usepackage{booktabs,array}
\usepackage{tikz}
\usetikzlibrary{arrows.meta}
\usepackage[font=small,labelfont=bf]{caption}
\usepackage[final,expansion=false]{microtype}
\usepackage[hidelinks]{hyperref}
\newtheorem{theorem}{Theorem}
\newcommand{\rootstep}{\mathrel{\xrightarrow{\epsilon}}}
\newcommand{\parentstep}{\mathrel{\xrightarrow{\rho}}}
\newcommand{\D}{\Downarrow_A}
\newcommand{\id}{\mathrm{id}_A}
\newcommand{\CT}{\operatorname{CT}_A}
\newcommand{\NF}{\mathrm{NF}}
\newcommand{\fib}[1]{F_{#1}}
\title{A counterexample to Lemma 64\\of the Kahrs and Smith proof for RTA-79}
\author{Moses Rahnama\\\small Mina Analytics}
\date{}
\hypersetup{pdftitle={A counterexample to Lemma 64 of the Kahrs and Smith proof for RTA-79},pdfauthor={Moses Rahnama}}
\begin{document}
\maketitle
\begin{abstract}
A complete targeted proof graph need not be universal, even when $\Downarrow_A$ is a consistency invariant. This gives a counterexample to Lemma~64 of Kahrs and Smith, \emph{Non-$\omega$-Overlapping TRSs are UN} (FSCD~2016). For the rule $d(x)\to x$, the carrier $\{c,d(c),d(d(c))\}$ admits such a graph with the sole parent edge $d(c)\mapsto d(d(c))$. Its represented equivalence omits the root equation $d(c)\Downarrow_A c$. The proof checks the consistency-invariant premise and every graph condition, and identifies the failed induction step in the published argument. Changing a parent edge gives a universal graph on the same carrier. Problem~79 of the RTA list of open problems is the conjecture that a term rewriting system has unique normal forms if its rules do not overlap under unification with infinite terms. The published solution of \mbox{RTA-79} depends on Lemma~64. Therefore, the published proof is false, and the \mbox{RTA-79} problem remains open.
\end{abstract}

\section{The assertion and its definitions}

Kahrs and Smith~\cite{KS} use finite proof graphs to establish transitivity of a relation $\Downarrow$ in their proof of unique normal forms for non-$\omega$-overlapping term rewriting systems, published as a solution of Problem~79 of the RTA list of open problems~\cite{RTA79}. A proof graph is a forest of directed parent edges. Two nodes represent the same class when their parent paths meet. Their Lemma~64, on page~22:15, asserts that, whenever $\Downarrow_A$ is a consistency invariant, every complete targeted proof graph on $A$ represents all of $\Downarrow_A$.

An extension keeps every existing parent edge (Definition~49 of~\cite{KS}). A graph is \emph{complete} when no further parent edge can be added while retaining the graph conditions. The example has one permitted edge opposite to a rewrite step. Keeping that edge prevents the graph from recording another rewrite equation. All relations and intermediate nodes are restricted to $A$.

\paragraph{Carriers and lifted relations.}
A \emph{strongly finite term-coalgebra} is a finite set $A$ of finite terms closed under subterms~\cite[Definition~5]{KS}. The signature is split into constructors and destructors. Frozen variables, when present, count as nullary constructors. For a relation $E$ on $A$, the destructor lifting $\overline E$ relates
$F(a_1,\ldots,a_m)$ and $F(b_1,\ldots,b_m)$ when $F$ is a destructor and $a_i\,E\,b_i$ for every $i$. The constructor lifting $\widehat E$ is defined in the same way for constructors. Put $\widetilde E=\overline E\cup\widehat E$. A relation is \emph{signature-closed} when $\widetilde E\subseteq E$; $\CT(E)$ is its least signature-closed extension~\cite[Definitions~24 and~25]{KS}.

\paragraph{The invariant.}
Write $\rootstep_A$ for one root contraction with both endpoints in $A$, $E^{-1}$ for the converse relation, and $E\cdot H$ for composition in path order. Definitions~38 and~40 of~\cite{KS} give
\begin{equation}\label{eq:fixedpoint}
 \D=\mu X.\bigl(X^{-1}\cup((\rootstep_A\!\cup\overline X)\cdot X)
                  \cup\id\cup\widehat X\cup X\bigr).
\end{equation}
Here $\mu$ denotes the least fixed point and $\id$ the identity relation. Thus $\D$ contains identity, is closed under converse and the two lifted operations, and is closed under prefixing by a root contraction or a destructor-lifting step. Formula~\eqref{eq:fixedpoint} has no transitivity clause.

A \emph{consistency invariant} is a consistent, signature-closed relation $S$ such that, for every constructor-compatible equivalence $E\subseteq S$,
\begin{equation}\label{eq:consistency}
 (\rootstep_A)^{-1}\cdot\overline E\cdot(\rootstep_A)
 \ \subseteq\ \CT(E).
\end{equation}
Consistency forbids relating distinct frozen variables. Constructor compatibility requires related constructor-topped terms to have the same head and related arguments~\cite[Definitions~27 and~30]{KS}. The middle factor in~\eqref{eq:consistency} is the destructor lifting $\overline E$, as typeset in Definition~30 of~\cite{KS}.

\paragraph{Proof graphs.}
Write $H^*$ for the reflexive-transitive closure of a relation $H$ and $\NF_\rho$ for the nodes with no parent. For a parent relation $\parentstep$ on $A$, put
$E_\rho=(\parentstep\cup{\xleftarrow{\rho}})^*$. It is a \emph{proof graph} when each node has at most one parent, every parent path terminates, and
\begin{equation}\label{eq:graph}
 E_\rho\subseteq\D,
 \qquad
 \parentstep\ \subseteq\ \rootstep_A\cup\overline{\D}\cup\widehat{E_\rho}.
\end{equation}
The union on the right is the \emph{grey edge relation}~\cite[Definitions~46 and~51]{KS}. Definitions~46, 51, 57 and~59 of~\cite{KS} use the destructor lifting $\overline{\D}$; it gives constructor-topped nodes singleton fibers, as noted after Definition~57. A proof graph $\beta$ extends $\rho$ when $\parentstep\subseteq{\xrightarrow{\beta}}$. Completeness means there is no proper such extension; universality means $E_\rho=\D$~\cite[Definitions~49, 54, and~63]{KS}. Determinism and termination make the equivalence-closure and common-descendant descriptions of $E_\rho$ agree.

\paragraph{Targets.}
For $t\in A$, put $\fib t=\{u\in A:t\,(\overline{\D})^*u\}$. These are the fibers called $E_t$ in~\cite[Definition~57]{KS}; the letter $F$ here distinguishes them from graph equivalence. A target map $q:A\to A$ is constant on each fiber, selects a member of it, and selects a root redex with contractum in $A$ whenever the fiber contains one. A proof graph with this target is \emph{targeted} when, for every $t,u\in A$,
\begin{align}
 &t\,(\overline{\D}\cap\parentstep)^*q(t)
       \quad\text{or}\quad t\in\NF_\rho,\label{eq:targetpath}\\
 &q(t)\parentstep u\quad\Longrightarrow\quad u\notin\fib t.
       \label{eq:targetexit}
\end{align}
These are Definitions~58 and~59 of~\cite{KS}, with the target on the quotient written as a fiber-constant map on $A$.

\section{The three-term counterexample}

Let $c$ be a nullary constructor, $d$ a unary destructor, and take
\begin{equation}\label{eq:terms}
 R=\{d(x)\to x\},\qquad
 t_0=c,\quad t_1=d(c),\quad t_2=d(d(c)),\qquad
 A=\{t_0,t_1,t_2\}.
\end{equation}
The carrier is strongly finite. The rule satisfies the right-hand-side variable condition and is a constructor-system rule. Its only proper left-hand-side subterm is a variable, so it has no nontrivial overlap, even with infinite substitutions. It is therefore strongly almost non-$\omega$-overlapping. Corollary~45 of~\cite{KS} gives that $\D$ is a consistency invariant on every term-coalgebra.

The same example occurs inside the constructor translation~\cite[Definition~12]{KS} of the system $D(x)\to x$ over $\{C,D\}$, with $C$ nullary and $D$ unary. That translation has the four symbols $C_c,C_d,D_c,D_d$ and the sole rule $D_d(x)\to x$. Set $c=C_c$ and $d=D_d$; the other two symbols occur in no node of $A$ and add no edge to its restricted relations. The carrier $A$ is the subterm closure of the conversion $d(d(c))\to d(c)\to c$, as in the construction in the proof of Theorem~66 of~\cite{KS}.

Define the parent relation and targets by
\begin{equation}\label{eq:example}
 \parentstep\ =\{(t_1,t_2)\},\qquad
 q(t_0)=t_0,\qquad q(t_1)=q(t_2)=t_2.
\end{equation}
The parent map is undefined at $t_0$ and $t_2$.

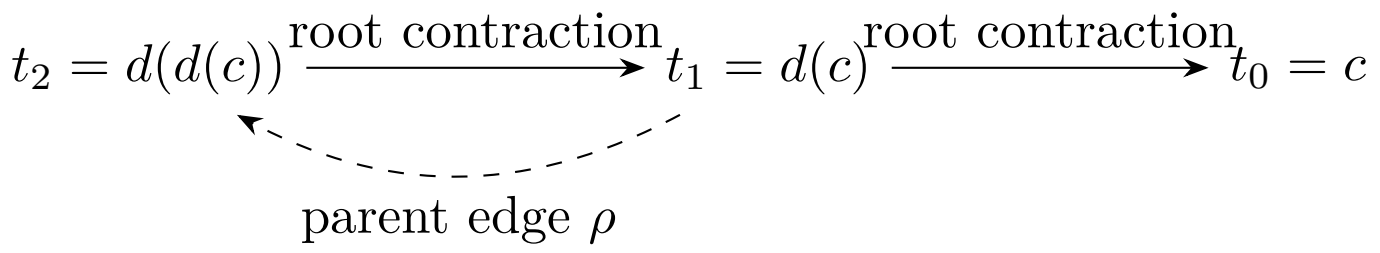
\begin{figure}[ht]
\centering
\begin{tikzpicture}[>=Stealth, every node/.style={font=\small}]
\node (t2) at (0,0) {$t_2=d(d(c))$};
\node (t1) at (4.1,0) {$t_1=d(c)$};
\node (t0) at (7.6,0) {$t_0=c$};
\draw[->] (t2) to node[above] {root contraction} (t1);
\draw[->] (t1) to node[above] {root contraction} (t0);
\draw[->,dashed,bend left=28] (t1) to node[below] {parent edge $\rho$} (t2);
\end{tikzpicture}
\caption{Solid arrows mark the two root contractions in $A$, and the dashed arrow marks the sole parent edge of $\rho$. The equivalence $E_\rho$ has classes $\{t_0\}$ and $\{t_1,t_2\}$.}
\label{fig:counterexample}
\end{figure}

\begin{theorem}\label{thm:counterexample}
For~\eqref{eq:terms}, $\D$ is a consistency invariant, and~\eqref{eq:example} is a complete targeted proof graph that is not universal. Lemma~64 of~\cite{KS} is therefore false as stated.
\end{theorem}

\begin{proof}
\smallskip\noindent\textit{The invariant and its fibers.}
The restricted root relation consists of $t_2\rootstep t_1$ and $t_1\rootstep t_0$. Identity and root prefixing in~\eqref{eq:fixedpoint} give
$t_1\D t_0$ and $t_2\D t_1$. Prefixing the first of these by $t_2\rootstep t_1$ gives $t_2\D t_0$. Together with their converses and the three diagonal pairs, these are all nine pairs in $A^2$. Hence
\begin{equation}\label{eq:allpairs}
 \D=A^2,\qquad
 \fib{t_0}=\{t_0\},\qquad
 \fib{t_1}=\fib{t_2}=\{t_1,t_2\}.
\end{equation}
\smallskip\noindent\textit{The consistency-invariant premise.}
The relation $A^2$ is signature-closed relative to $A$. It is consistent in the sense of~\cite{KS} because $A$ contains no frozen-variable nodes; it is also constructor-compatible, since $c$ is its only constructor-topped node. To check~\eqref{eq:consistency}, suppose
\[
 u\mathrel{\xleftarrow{\epsilon}}a\;\overline E\;b\rootstep v.
\]
The sole rule forces $a=d(u)$ and $b=d(v)$. The middle premise then says $u\,E\,v$, hence $u\,\CT(E)\,v$. This holds for every relation $E$ on $A$, in particular for every constructor-compatible equivalence.

\smallskip\noindent\textit{The proof-graph and targeting conditions.}
The parent relation is deterministic, and every parent path has length at most one. Its represented equivalence is
\begin{equation}\label{eq:classes}
 E_\rho=\{(t_0,t_0)\}\ \cup\ \{t_1,t_2\}^2,
\end{equation}
and this relation is contained in $\D$. The parent edge $t_1\parentstep t_2$ belongs to $\overline{\D}$: its two arguments are $t_0$ and $t_1$, and $t_0\D t_1$. Thus~\eqref{eq:graph} holds, although this parent edge is not a rewrite step.

The map $q$ is fiber-constant by~\eqref{eq:allpairs}. Both destructor nodes are root redexes with contracta in $A$, and their chosen target $t_2$ is one of them. Node $t_1$ reaches $t_2$ by its single parent edge in $\overline{\D}$; $t_0$ and $t_2$ reach their own targets in zero steps. This proves~\eqref{eq:targetpath}. Both selected targets are parent normal forms, so~\eqref{eq:targetexit} holds as well.

\smallskip\noindent\textit{Completeness.}
Let $\beta$ be any proof-graph extension of $\rho$. It must retain $t_1\xrightarrow{\beta}t_2$. Table~\ref{tab:extensions} excludes every possible additional edge. The exclusions apply to the grey relation recomputed from $\beta$.

\begin{table}[ht]
\centering\small
\begin{tabular}{@{}p{0.25\linewidth}p{0.69\linewidth}@{}}
\toprule
Additional edge & Reason it is impossible in a proof graph extending $\rho$\\
\midrule
$t_1\to t_0$ or $t_1\to t_1$ & Conflicts with the retained parent $t_1\to t_2$ and determinism.\\[3pt]
$t_2\to t_1$ & Forms a directed cycle with $t_1\to t_2$.\\[3pt]
$t_2\to t_2$ or $t_0\to t_0$ & Is a self-loop, contrary to termination.\\[3pt]
$t_2\to t_0$ & Is not one root contraction; the source and target have different head types, so neither lifted relation applies.\\[3pt]
$t_0\to t_1$ or $t_0\to t_2$ & Has no root rule and belongs to neither lifted relation.\\
\bottomrule
\end{tabular}
\caption{Each of the eight ordered pairs other than the parent edge fails a condition on extensions, so $\rho$ is complete.}
\label{tab:extensions}
\end{table}

The path $t_2\rootstep t_1\rootstep t_0$ has two root contractions, whereas a grey root edge is a single contraction. Thus it gives no parent edge $t_2\to t_0$. Recomputing $E_\beta$ cannot change the head restrictions of $\widehat{E_\beta}$. No additional edge is possible, so $\beta=\rho$. This proves completeness among all proof graphs, hence also among targeted ones.

\smallskip\noindent\textit{Non-universality.}
The root equation $t_1\D t_0$ holds, but~\eqref{eq:classes} gives $\neg(t_1\,E_\rho\,t_0)$. Therefore $E_\rho\ne\D$.
\end{proof}

\section{The failed inference in the published proof}

The proof of Lemma~64 reduces universality to representing every root contraction. It then compares a root redex with the selected redex in its destructor fiber and invokes the consistency-invariant property. The consistency condition requires their arguments to be related by $E_\rho$, as expressed by the bar in~\eqref{eq:consistency}. The published proof appeals to induction to obtain this barred relation from the fiber path $t\,(\overline{\D})^*q(t)$.

For the root step $t_1\rootstep t_0$, the selected redex is $q(t_1)=t_2$, with root contraction $t_2\rootstep t_1$. The two relevant comparisons are
\begin{align*}
 t_1\;\overline{\D}\;t_2
   &\quad\Longleftrightarrow\quad t_0\D t_1,
       &&\text{both sides hold},\\
 t_1\;\overline{E_\rho}\;t_2
   &\quad\Longleftrightarrow\quad t_0\,E_\rho\,t_1,
       &&\text{both sides fail}.
\end{align*}
The parent edge gives $t_1\,E_\rho\,t_2$. It also represents the selected target's root equation $t_2\,E_\rho\,t_1$. Neither fact gives the second displayed comparison. Constructor compatibility permits decomposition at constructor heads; these two heads are destructors.

The use of~\eqref{eq:consistency} at this step requires the fork
\[
 t_0\mathrel{\xleftarrow{\epsilon}}t_1
       \;\overline{E_\rho}\;t_2\rootstep t_1.
\]
Its middle premise is absent. Replacing it by $t_1\,E_\rho\,t_2$ does not satisfy Definition~30. The induction hypothesis invoked at this step is the claim for the argument pair $(t_0,t_1)$. The pair is $\D$-related through the root step $t_1\rootstep t_0$, the case under consideration, so the induction returns to the pair it started from, and the graph does not represent it.

The conclusion of Lemma~62 holds here: $E_\rho$ is signature-closed on $A$, since $\widehat{E_\rho}=\{(t_0,t_0)\}$ and $\overline{E_\rho}=\{(t_1,t_1),(t_2,t_2)\}$ are both contained in $E_\rho$. Closure under adding a head gives no rule for removing a destructor head.
\par Two later workshop papers by the same authors concern this step. The short version~\cite{KSIWC} describes the inference, the assumption on subterms that the graph equivalence coincides with the invariant, as a ``somewhat problematic fixpoint argument'' and sketches a coinductive variant. Kahrs's extended abstract~\cite{KahrsIWC} replaces the invariant and builds the proof graph from the root rewrite edges; it states universality with an outline of an induction on pairs of terms.
\section{Repair on the same carrier}

Choose the lower target $q'(t_1)=q'(t_2)=t_1$, keep $q'(t_0)=t_0$, and use the parent chain
\[
 t_2\xrightarrow{\rho'}t_1\xrightarrow{\rho'}t_0.
\]
Both edges are root contractions. Every node reaches $t_0$, so $E_{\rho'}=A^2=\D$. The graph is deterministic and terminating. It is targeted: $t_2$ reaches its selected target $t_1$ inside the destructor fiber, and the parent edge leaving that target goes to $t_0$, outside the fiber. It is complete because the parents of $t_1,t_2$ are fixed and no permitted terminating edge leaves $t_0$.

This repair retains every equation represented by $\rho$ but replaces its parent edge $t_1\mapsto t_2$ by $t_1\mapsto t_0$. It is therefore not an extension in Definition~49.

The proof of Lemma~65 starts from the empty graph with an arbitrary target and completes it through Lemma~61. With the target $t_2$, the extension step of Lemma~61 adds the edge $t_1\mapsto t_2$ along the chain $t_1\,\overline{\D}\,t_2$, and the result is $\rho$, which is complete. Thus arbitrary target selection followed by edge completion can end in a nonuniversal graph. Lemma~65 is proved from Lemma~64 (its printed proof cites Lemma~61 and, by a misprint, Lemma~65 itself), Theorem~66 from Lemma~65, and Corollary~67 and Theorems~68 and~69 from Theorem~66, so the printed proofs of all five fail with Lemma~64.


\begin{thebibliography}{1}
\bibitem{KS}
Stefan Kahrs and Connor Smith.
Non-$\omega$-Overlapping TRSs are UN.
In Delia Kesner and Brigitte Pientka, editors,
\emph{1st International Conference on Formal Structures for Computation and Deduction (FSCD 2016)},
volume~52 of \emph{Leibniz International Proceedings in Informatics},
pages~22:1\textendash22:17. Schloss Dagstuhl, 2016.
\href{https://doi.org/10.4230/LIPIcs.FSCD.2016.22}{doi:10.4230/LIPIcs.FSCD.2016.22}.

\bibitem{RTA79}
Nachum Dershowitz, Jean-Pierre Jouannaud, and Jan Willem Klop.
Problems in rewriting III.
In Jieh Hsiang, editor,
\emph{Rewriting Techniques and Applications (RTA 1995)},
volume~914 of \emph{Lecture Notes in Computer Science},
pages~457\textendash471. Springer, 1995.
Problem~79, originator Mizuhito Ogawa; current entry at
\url{https://www.cs.tau.ac.il/~nachum/rtaloop/problems/79.html}.

\bibitem{KSIWC}
Stefan Kahrs and Connor Smith.
Non-omega-overlapping TRSs are UN.
In Beniamino Accattoli and Ashish Tiwari, editors,
\emph{Proceedings of the 5th International Workshop on Confluence (IWC 2016)},
pages~11\textendash15. Obergurgl, 2016.
\url{https://www.csl.sri.com/users/tiwari/iwc2016/iwc2016.pdf}.

\bibitem{KahrsIWC}
Stefan Kahrs.
Consistency and Invariants (Extended Abstract).
In Beniamino Accattoli and Bertram Felgenhauer, editors,
\emph{Proceedings of the 6th International Workshop on Confluence (IWC 2017)},
pages~1\textendash6. Oxford, 2017.
\url{https://cl-informatik.uibk.ac.at/images/events/IWC/iwc2017.pdf}.
\end{thebibliography}
\end{document}